\documentclass{mn2e}
\usepackage[]{color}
\usepackage[colorlinks,citecolor=blue]{hyperref}
\usepackage[]{graphicx}

\usepackage[]{natbib}
\usepackage[]{lscape}
\usepackage[twoside]{rotating}
\usepackage[]{enumerate}
\usepackage[]{lettrine}
\usepackage[]{times}
\usepackage[]{astroj}
\title[Chemically tagging the Solar neighborhood ]{Quantitative chemical tagging, stellar ages and the chemo- dynamical evolution of the Galactic disc }
\author[Mitschang et al. ]{\parbox{\textwidth}{A. W. Mitschang$^{1,2}$ \thanks{E-mail: arik.mitschang@mq.edu.au}, G. De Silva$^{3}$, D. B. Zucker$^{1,2,3}$, B. Anguiano$^{1,2}$, T. Bensby$^{4}$, S. Feltzing$^{4}$}
\vspace{0.3em}
\\
$^{1}$  Macquarie University Research Centre in Astronomy, Astrophysics \& Astrophotonics, NSW 2109, Australia \\
$^{2}$  Department of Physics \& Astronomy, Macquarie University, NSW 2109, Australia \\
$^{3}$  Australian Astronomical Observatory, PO Box 296, NSW 1710, Australia \\
$^{4}$  Lund Observatory, Department of Astronomy and Theoretical Physics, Box 43, SE-221 00 Lund, Sweden \\
}
\begin{document}
\maketitle
\begin{abstract}
The early science results from the new generation of high-resolution
stellar spectroscopic surveys, such as GALAH and the Gaia-ESO survey,
will represent major milestones in the quest to chemically tag the
Galaxy. Yet this technique to reconstruct dispersed coeval stellar
groups has remained largely untested until recently. We build on
previous work that developed an empirical chemical tagging probability
function, which describes the likelihood that two field stars are
conatal, that is, they were formed in the same cluster environment. In
this work we perform the first ever blind chemical tagging experiment,
i.e., tagging stars with no known or otherwise discernable
associations, on a sample of 714 disc field stars with a number of
high quality high resolution homogeneous metal abundance
measurements. We present evidence that chemical tagging of field stars
does identify coeval groups of stars, yet these groups may not
represent distinct formation sites, e.g. as in dissolved open
clusters, as previously thought. Our results point to several
important conclusions, among them that group finding will be limited
strictly to chemical abundance space, e.g. stellar ages, kinematics,
colors, temperature and surface gravity do not enhance the
detectability of groups. We also demonstrate that in addition to its
role in probing the chemical enrichment and kinematic history of the
Galactic disc, chemical tagging represents a powerful new stellar age
determination technique.
\end{abstract}
\section{Introduction}
\label{introduction}
The field of Galactic Archaeology, aimed at uncovering the events that
led to the current state of the Milky Way -- and more broadly to
spiral galaxies in general -- harnesses the unique observational
property of our own Galaxy: that we can resolve individual
stars. Though many large photometric surveys have taken advantage of
this, to date there have been few large scale spectroscopic surveys
observing Milky Way stars. Notably, the SEGUE (\citealt{yanny2009}) and
RAVE (\citealt{steinmetz2006}) surveys have enabled important advancements
in the understanding of the dynamical nature of the Galaxy. Both
surveys were done at low resolution; high resolution counterparts at
the same scale have yet to come. However, this is set to change in the
coming years, as the Gaia-ESO public spectroscopic survey
(\citealt{gilmore2012}) continues observations of upwards of 100 000 stars,
and the unprecedented million star survey, GALAH (Galactic Archaeology
with HERMES; \citealt{freeman2013}) begins operations at the
Anglo-Australian Astronomical Telescope (AAT) in late 2013. Combined
with the precise astrometry of the Gaia space telescope mission
\footnote{http://sci.esa.int/gaia}, we will soon have an extremely detailed
and comprehensive picture of millions of Galactic stars.

Of great importance to the study of the evolution of the Galaxy as a
whole is the chemical and kinematic evolution of the disc. The disc is
where most star formation occurs, it is rich with astrophysical
fossils and is relatively easy to observe (compared to the stellar
halo or bulge/bar). A common view of the physical structure of the
Galaxy is that there are two major components of the disc: a thick,
diffuse disc with a scale height of order 1 kiloparsec, and a compact
and dense thin component with a scale height of about 300 parsecs
(\citealt{gilmore1983}). In this paradigm, the thick disc stars are old,
metal poor, and have large dispersions in their vertical space
motions. The thin disc, on the other hand, is young, metal rich, and
has a small vertical velocity dispersion
(\citealt{bensby2003,nordstroem2004,bensby2005,anguiano2013a}). A
healthy debate continues as to the origin of the thick disc. Is it a
product of a galactic collision or tidal interactions with dwarf
galaxies (\citealt{quinn1993,abadi2003})? Or does the process of stellar
radial migration
(\citealt{sellwood2002,minchev2011,loebman2011,roskar2013}) play a dominant
role in the kinematic heating? Perhaps none of these explains the
existence of the thick disc. Another paradigm for describing the
Galactic disc as a whole is that, instead of two monolithic components
with distinct evolutionary paths, there is a smooth distribution of
``mono-abundance populations'' (MAPs) which rise out of constant heating
and star formation cycles (\citealt{bovy2012,rix2013}). To address such
questions requires detailed chemical and kinematic analyses of large
numbers of stars. Without a doubt, Gaia will accomplish the
latter. The former is where GALAH will make great strides.

GALAH is not just a high-resolution spectroscopic survey of a million
stars. A primary mission of the survey is to ``chemically tag'' the
entire sample in order to search for long-since-dispersed star
clusters. \citet{fbh2002} introduced the concept of chemical tagging, in
which stars are linked to individual star formation events when their
abundance patterns in a range of elements, from $\alpha$ to Fe-peak,
light to heavy s and r-process, are the same. This is possible, in
theory, because star formation within clusters occurs in rapid bursts
within a giant molecular cloud which is well mixed with enriched
material from a previous generation of stars. It is thought that all
stars in the Galaxy are formed within such clusters and disperse on
time-scales of typically tens to several hundred Myr in the Solar
neighborhood (\citealt{janes1988,lada2003}). The chemically tagged stars
then would be considered \emph{conatal}, or having formed in the same
molecular cloud, localised within the Galactic disc, implying also
that they are \emph{coeval}, having formed at the same epoch.

For chemical tagging to work in practice, it is important that star
clusters that represent typical star formation (insofar as possible)
be homogeneous in their abundance patterns. Open clusters and moving
groups have been shown to exhibit uniform abundance patterns based on
high resolution, high signal-to-noise abundance analyses
(\citealt{desilva2006,desilva2007b,desilva2007a,bubar2010,pancino2010}). In addition
to the requirement of homogeneity within open clusters, chemical
tagging relies on the adequate differentiation of distinct clusters in
abundance space. Recently, \citet{mitschang2013} used a high resolution
spectroscopic sample of Galactic open cluster stars from the
literature to quantify the level to which chemical tagging can
distinguish between conatal and disparate stars. They developed a
chemical difference metric, $\delta$$_\mathrm{C}$, which decomposes the
N-dimensional chemical abundance space. An empirical probability
function was derived, which allows confident tagging of pairs of stars
using the $\delta$$_\mathrm{C}$ metric. The dimensionality of chemical tagging
abundance space has also been probed. \citet{ting2012} used a principal
component abundance analysis (PCAA) to discover the elements with the
largest global variance, finding 8-9 elements form a truly independent
set. These will be the most powerful chemical tags and will be the
target of chemical tagging surveys.

There have been several studies applying the concepts of chemical
tagging on small scales. \citet{desilva2011} and \citet{tabernero2012} used
chemical information to make membership decisions on stars compatible
with the Hyades supercluster association, as did \citet{pompeia2011} with
the Hyades stream to test its origin. \citet{desilva2013} found that the
Argus association stars probably originated from the open cluster IC
2391, using chemical tagging supported by kinematic and chronological
information. Conversely, \citet{carretta2012} found chemical information
from a handful of elements indicated several distinct populations in
the open cluster NGC 6752. Beyond the Galaxy, but still within reach
of current instrumental capabilities, even coeval groups in a dwarf
galaxy have been tagged using chemical information
(\citealt{karlsson2012}). It is important to note that these studies were
able to make qualitative decisions based on chemical signatures due to
\emph{a priori} information on their likelihood of membership in an
association. There will be no such advantage for a large scale
chemical tagging experiment looking for dispersed coeval
structures. To date, no \emph{blind} chemical tagging experiment, one where
the sample stars have no known associations, has been carried out; this is
the goal of the current work.

In the context of Galactic evolution, accurate ages must play an
integral role. Stellar ages, however, are notoriously difficult to
infer for single stars (see \citealt{soderblom2010} for a comprehensive
review). The most common method, isochrone fitting, has significant
limitations, and although astroseismology can produce very accurate
ages, it takes significant observational investment and may not yet be
appropriate for all stars (\citealt{soderblom2010}). An important point is
that, aside from the Sun, the most accurate ages, and those which set
the basis for statistical age relations and stellar model
calibrations, are those from open clusters (conatal and coeval stellar
groups).

In this paper we aim to characterise chemical tagging via the group
finding technique from \citet{mitschang2013}, conducting the first blind
chemical tagging experiment at any scale. We show that results from
the chemical tagging are consistent with appropriate coeval linking
based on several indicators, but that information classically used as
membership indicators in conatal groups (open clusters), e.g. velocity
dispersions, photometric colors and magnitudes, surface gravity and
effective temperatures, etc., will not typically aid in identifying
contaminated groupings. The \emph{predictions} section in
\citet{soderblom2010} does not mention chemical tagging as a future
source of stellar age determination, yet we argue here that this is
amongst the most important outcomes these experiments will yield. That
is, instead of independently measured stellar ages being used to
scrutinize the tagging of coeval groups, rather chemical tagging will
produce more accurate and precise ages for a larger number of stars
than otherwise would be possible.

\section{Disc field chemical abundance data}
\label{disc field chemical abundance data}
We perform our chemical tagging experiment on a large sample of 714
disc field dwarfs and subgiants, observed using the high-resolution
spectrographs FEROS at the ESO 1.5m and ESO 2.2m telescopes, SOFIN at
the NOT, MIKE at the Magellan Clay telescope, and UVES at the ESO VLT
telescope. Typical resolutions of these observations range between
R=42,000 to 110,000, and signal-to-noise ratios are typically
above 250.  The analysis of the spectra, including computing stellar
parameters and element abundances is described in \citet{bensby2013}.

This homogeneous abundance sample of mostly dwarf and turn-off stars
inhabits a volume with an approximately 150 parsec radius centred on
the Sun. All 714 stars have trigonometric parallaxes and proper
motions taken from the TYCHO-2 catalogue (\citealt{hog2000}), so in
addition to the 12 dimensional chemical abundance space comprised of
Fe, Na, Mg, Al, Si, Ca, Ti, Cr, Ni, Zn, Y, Ba, we also have 6
dimensions of position and kinematic space. Ages for individual stars
in this sample have been derived using an isochrone fitting method
described in \citet{bensby2011}. The high precision in the analysis and
chemical dimensionality (see, e.g., \citealt{ting2012,mitschang2013}) along
with the size of this sample make it an ideal data-set for the first
truly blind chemical tagging experiment.

\section{Chemical Tagging}
\label{chemical tagging}

\subsection{Group finding procedure}
\label{group finding procedure}
We performed group finding using the algorithm described in
\citet{mitschang2013}. The procedure works by computing the pairwise
metric $\delta$$_\mathrm{C}$ defined as

\[
\delta_C=\sum_{C}^{N_C}\omega_C\frac{|A_{C}^i-A_{C}^j|}{N_C}
\]

where N$_\mathrm{C}$ is the number of measured abundances and A$^\mathrm{i}$$_\mathrm{C}$ and A$^\mathrm{j}$$_\mathrm{}$C
are individual abundances of element C for stars i and j,
respectively. The $\omega$$_\mathrm{C}$ factor may be used to give more or less
weight to a particular element, with respect to the overall chemical
difference, given some external knowledge about that element. In this
case, given the lack of detailed study into this factor, we fix it at
unity for all elements involved in the $\delta$$_\mathrm{C}$ computation. This
metric is then computed over all pairs of stars, describing the
difference in abundance patterns of those particular stars.

Then, utilizing the empirical probability function based on open
clusters derived in \citet{mitschang2013}, the $\delta$$_\mathrm{C}$ values are
translated to a probability, $P_{lim}$ that a given pair i and j are
conatal (i.e. formed in open cluster-like star formation
events). Simply understanding if two stars have a reasonable chance of
being conatal is interesting, but we seek to find groups that have a
high \emph{combined} probability, essentially reassembling a long-since
dispersed cluster of stars.

Our algorithm for linking groups begins with the highest density
clustering in $\delta$$_\mathrm{C}$ space and assembles groups such that all star
pairs meet the required probability threshold. Briefly, the algorithm
proceeds as follows: all pairs for which the probability is less than
our threshold are first removed, and the remainder sorted by highest
probability first. Chemically tagged groups of stars are formed by
linking pairs that share common stars such that an individual star
only inhabits one cluster. The sorting ensures only the best match for
pairs that may adequately match more than a single group. Linking then
proceeds down the chain of pairs, removing those pairs where one of
the stars has been assigned to a group. We did this for two values of
limiting probability, $P_{lim}$, of 90 and 68\%. The 68\% threshold,
corresponding to approximately a 1-$\sigma$ detection of the coeval
signature between two stars, is the lowest meaningful probability we
can tag with, yet at a $\delta$$_\mathrm{C}$ of 0.057 dex we are pushing the limits
in terms of abundance measurement uncertainties.

The sample used for calibration of the chemical tagging probability
function was selected from high-resolution studies with uncertainties
on the order of the current sample. Therefore, with respect to
internal errors, the function is appropriately applied to this study
at the 68\% level. The effects of systematic uncertainties due to the
heterogenous calibration sample are discussed in detail in
\citet{mitschang2013}; here we note that the 90\% threshold also explored
in the current work corresponds well to the 68\% threshold in the
simulated ``intrinsic'' probability function, which attempts to weigh
the contribution of external uncertainties. This can be understood,
loosely and in a global sense, as indicative of the errors on group
determination.

The results of group linking are summarised in Table
\ref{prob_grp_table}. For each threshold we list the corresponding limit
in $\delta$$_\mathrm{C}$ taken from the probability function ($\delta$$_\mathrm{C}$ lim), the
number of groups recovered for which there were at least three members
(Num), the mean $\delta$$_\mathrm{C}$$^\mathrm{avg}$ ($\delta$$_\mathrm{C}$$^\mathrm{avg}$ is the average of pair
probabilities over the recovered group), the mean and maximum number
of stars in the detections and the percentage of all stars in our sample
that were tagged to groups.

\begin{table}
\begin{center}
\caption{
The properties of groups recovered for various probability levels. 
\label{prob_grp_table}
}
\begin{tabular}{lrrrr}
$P_{lim}$&68\%&$68\%_{N_\star > 2}$&90\%&$90\%_{N_\star > 2}$\\
\hline
$\delta$$_\mathrm{C}$ lim (dex)&0.057&&0.033&\\
Num clusters&102&67&171&80\\
Mean $\delta$$_\mathrm{C}$$^\mathrm{avg}$ (dex)&0.039&0.037&0.025&0.024\\
Min $\delta$$_\mathrm{C}$$^\mathrm{avg}$ (dex)&0.026&0.026&0.011&0.015\\
Mean N$_\mathrm{\star{}}$&6.6&8.9&3.0&4.2\\
Max N$_\mathrm{\star{}}$&42&&12&\\
\% tagged&94&84&73&47\\
\end{tabular}
\end{center}
\end{table}
The link between stars and groups identified at the 68\% probability
level is given in Table \ref{stars_groups}, where each star identified
as part of a group with 3 or more members has its Hipparcos number
(HIP) listed along with a group identification number (GID).

\begin{table}
\begin{center}
\caption{Group linking between stars at 68\% probability detection
  threshold. The full version of this table is available with the
  arXiv source files (table\_2.dat). \label{stars_groups}}
\begin{tabular}{rl}
HIP&GID\\
\hline
80&6\\
305&13\\
407&1\\
699&18\\
768&19\\
$\vdots$&$\vdots$\\
\end{tabular}
\end{center}
\end{table}

\subsection{Results of group finding}
\label{results of group finding}
\begin{figure}
\centering
\includegraphics[width=1.0000\linewidth]{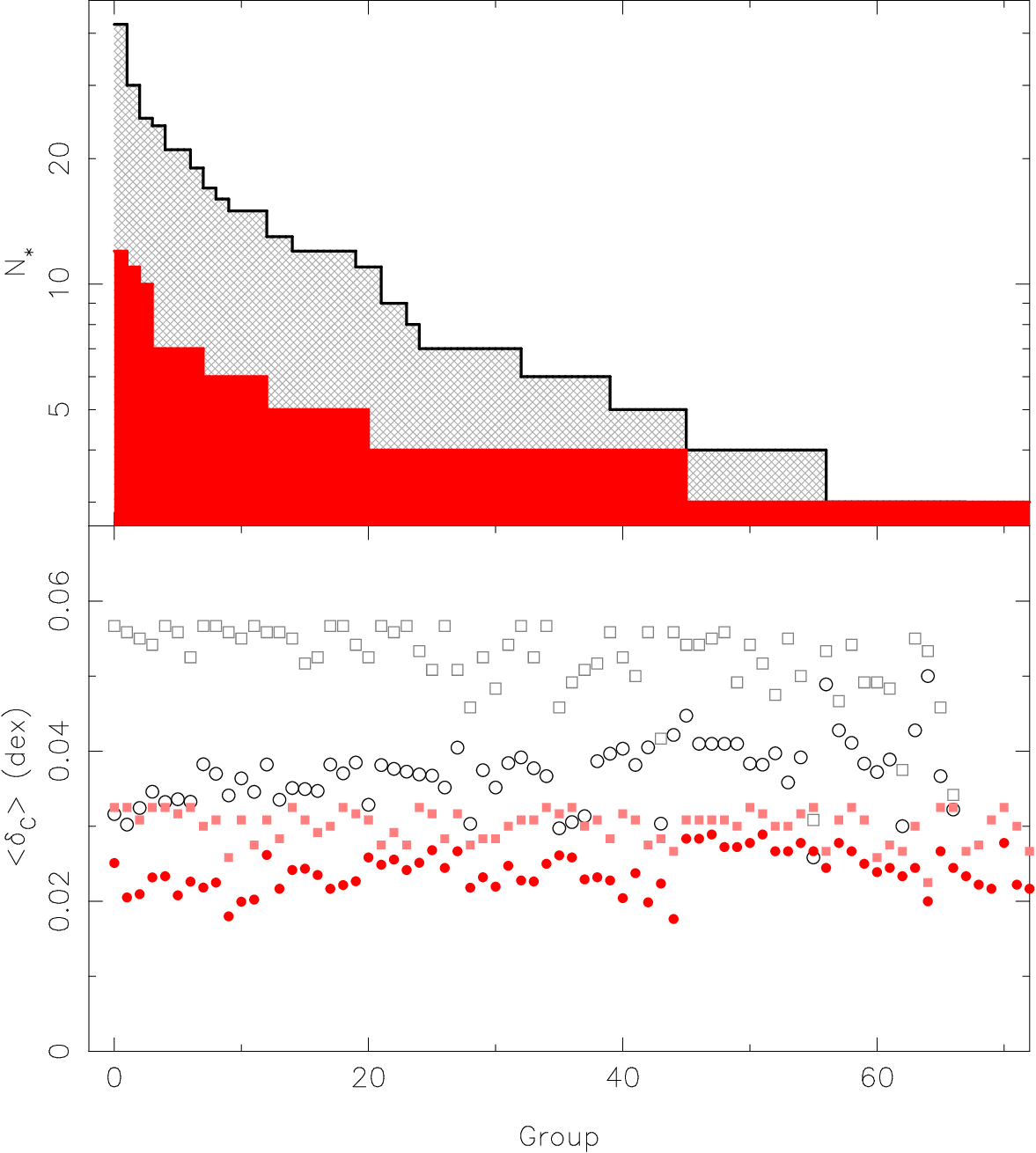}\\
\caption{The top panel shows the distribution of numbers of members
  for groups with 3 or more members, while the bottom panel shows the
  $\delta$$_\mathrm{C}$ values corresponding to the bins; squares are
  the max $\delta$$_\mathrm{C}$ and circles are the mean, in each
  group. Both panels are sorted by number of stars in the recovered
  group. In each panel $P_{lim}=90$\% is represented by red, while
  $P_{lim}=68$\% is black.
\label{grp_props}}
\end{figure}

The distribution of member counts per group and their respective
average $\delta$$_\mathrm{C}$ values are shown in Figure \ref{grp_props}. Note the
exponential drop off in the number of recovered members; a majority of
the groups have no more than 4 members. At $P_{lim}=68$\%, the largest
group has 38 members, and also exhibits one of the smallest mean
$\delta$$_\mathrm{C}$ values (or high mean pair probabilities). Indeed, we see a
slight rising trend in mean $\delta$$_\mathrm{C}$ with increasing member counts for
both probability thresholds in the lower panel of Figure
\ref{grp_props}. To some degree, this is explained by our algorithm's
preference for the highest density groups (those having smallest
$\delta$$_\mathrm{C}$), which are linked first.

\begin{figure*}
\centering
\includegraphics[width=0.2375\linewidth]{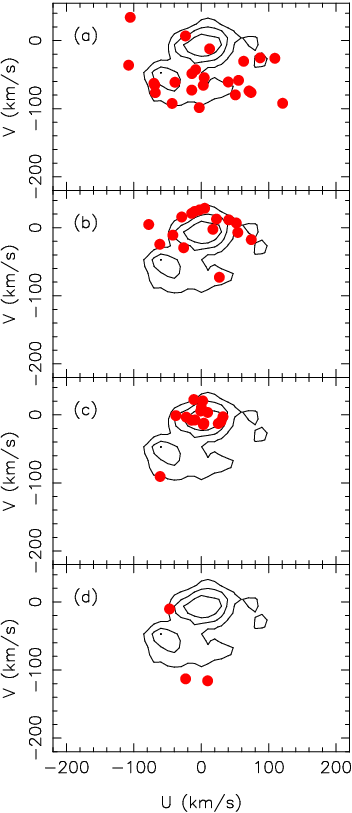}\includegraphics[width=0.2375\linewidth]{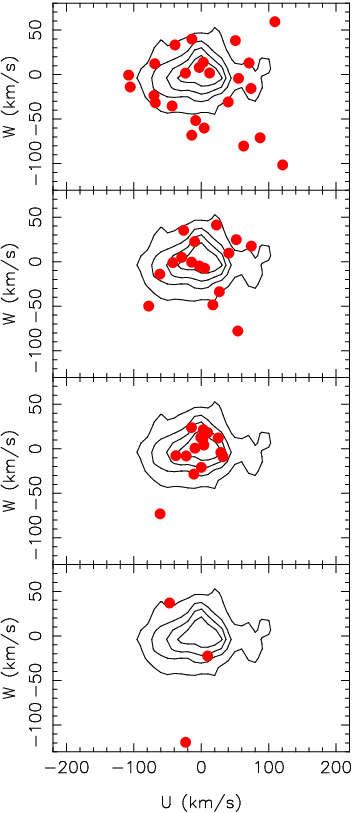}\includegraphics[width=0.2375\linewidth]{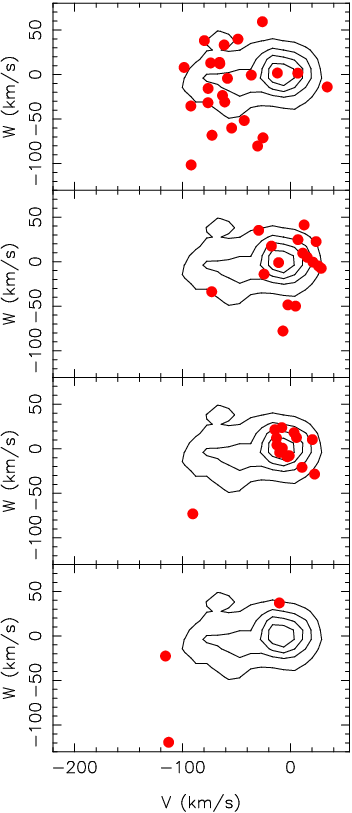}\includegraphics[width=0.2375\linewidth]{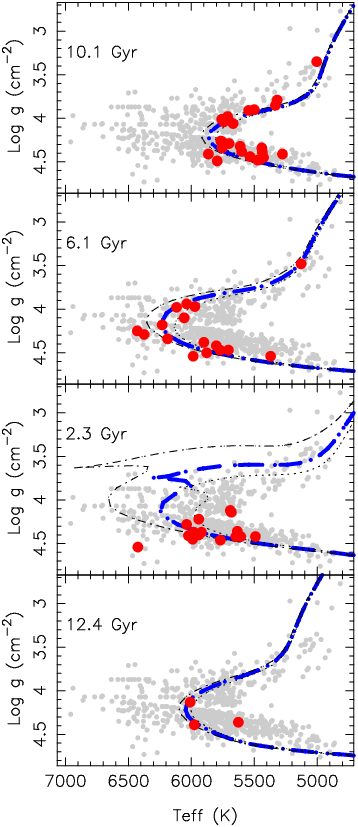}\\
\caption{UV, UW and VW kinematic plots for a sample of recovered groups along with T$_\mathrm{eff}$ and Log g CMD plots. The kinematic plots show the entire sample in black contours and members with red circles. The CMD plots show the entire sample in grey points, the member stars in red circles and the best-fitting isochrone to the group as a dashed blue line. The dot-dashed and dotted black lines are isochrones one Gyr younger and one Gyr older, respectively, for comparison. These groups were selected to illustrate the wide variety, and extremes, of kinematic and CMD orientations that are present in our recovered groups. 
\label{grp_examples}}
\end{figure*}

Amongst the recovered coeval groups there is a wide variety of
properties of the members in terms of stellar parameters, e.g. surface
gravity, effective temperature, space velocities, abundance patterns
and ages. There is over one dex of variation in the metallicities
between groups, yet all of them, by definition, have homogeneous
abundance patterns. Such diversity in a modest sample of stars as in
this study affords an excellent opportunity to not only study chemical
tagging, but also the science it will enable. Figure \ref{grp_examples}
shows four examples from the $P_{lim}=$ 68\% set showing orientations
in three kinematic planes (U, V, and W) and log g vs. T$_\mathrm{eff}$ planes,
with comparison to the entire sample, selected to illustrate some of
the range, and extremes, of recovered groups. Panel (a) shows one of
the largest tagged groups, which, though it exhibits large scatter in
kinematics, has a tight orientation in the CMD plane about its
best-fitting isochrone (see Section \ref{stellar age determinations}).
In panel (c), the main sequence of the group appears atypical by eye,
given the a seeming reverse slope, and is difficult to fit due to this
and the predominance of lower main-sequence dwarf stars, an issue
which affects many groups in the analysis (seen in panel d as well);
stars on the lower main-sequence provide only weak discriminatory
power between isochrones of different ages, due to the convergence of
evolutionary tracks at low surface gravities. However, it is important
to note that the atypical form may be an illusion; if the full
population of that group were available, it is possible the same
interpretation would not be made, perhaps save a single star. Panel
(d) shows a low membership group where mostly lower main-sequence
stars are identified, consequently making an age determined from it
less meaningful. Figure \ref{grp_examples} also highlights the range of
ages that the groups have from under one to 14 Gyrs.

Performing the calculation for total chemical tagging efficiency given
in \citet{mitschang2013}, based on their literature abundance sample, the
expected efficiency of chemical tagging at 68\% limiting probability
would be roughly 9\%, meaning that 9\% of the total sample of stars
could be reliably tagged. Combining the contamination rate of 50\% in
that study with the $\sim{}$80\% tagged in our experiment, we may have cleanly
tagged 40\% of the stars in our sample, which is significantly larger
than expectations. Moreover, the 9\% efficiency estimate was based
purely on the confluence of two separate star formation signatures in
chemical space, due to the fact that the abundance sample used
contained only stars from known open clusters. A field star sample
would be further complicated by dynamical mixing processes, which
means that the chemical tagging efficiency would have to be folded in
with the prior probability that any two random stars in a local
sample, regardless of their observed properties, are conatal. In this
context, the number of stars tagged seems at odds with the number of
conatal signatures we might expect. How likely is it that we would
find multiple (or any) such conatal associations in the Hipparcos
volume?

A comprehensive approach to answering that question would require
detailed modeling of Galactic evolution at the scale of individual
stars, tracking disrupting clusters over a large range of cosmic time,
possibly in the form of an N-body simulation. To our knowledge,
simulations of this nature have not been fully rendered yet.

In \citet{fbh2002} it is suggested that chemical tagging will probe
particular enrichment events, i.e., those which polluted a molecular
cloud resulting in a star formation episode discrete in space and
time. Those stars would then disperse around the Galaxy, retaining
their initial chemical patterns. Given the seeming implausibility of
detecting as many apparent coeval groups as we have, even for very
tight chemical differences, we offer several interpretations that may
explain our results:

\begin{enumerate}[1.)]
\item  The chemical overlap between conatal groups is far greater than
observed in \citet{mitschang2013}, resulting in high contamination,
and tagged groups represent nothing more than stars with similar
chemistry.
\item  Open clusters, or the current literature sample, do not adequately
represent typical star formation in the disc, resulting in the
contamination estimate being either too high or too low.
\item  Stellar dynamical mixing processes (e.g. radial migration,
churning) are not efficient, keeping members of unbound
associations in relative proximity.
\item  The star-formation and enrichment cycle, per epoch, is not
stochastic, yielding similar abundance patters as a function of age
to within current measurement abilities. In other words, chemically
tagged groups represent coeval, but \emph{not} conatal, stars.

\end{enumerate}

We will continue to discuss these interpretations of chemically
tagging groups for the remainder of this work. Initially, and for the
next section, involving stellar age determinations, we operate on the
traditional assumption of \citet{fbh2002}, that these groups represent
star formation sites similar to open clusters but which have
dispersed, i.e., they are considered conatal.

\section{Stellar Age determinations}
\label{stellar age determinations}
Perhaps one of the most powerful incentives to tag coeval groups of
stars is to enhance the reliability of determining ages for the stars
that make them up. There are many methods for determining ages for
single stars (\citealt{soderblom2010}), but by far the most common method
is by fitting isochrones to their positions on a CMD plane. The most
obvious difficulty here is that fitting any model curve to a single
point is a highly degenerate problem. With isochrones, this is
especially difficult in the lower main-sequence region, as tracks of
different ages converge with decreasing temperature on the main
sequence. Even in the turn-off region, where the track separation is
higher for any given difference in age, overlaps can resurface at very
large age separations.

More subtly, these fitting procedures aim to land the star exactly on
the isochrone. It is possible that scatter about model isochrones
exists beyond the measurement errors for open cluster stars. If it is
true that open clusters form quite rapidly (e.g., see \citealt{lada2003}),
and that a single isochrone describes the population well, then that
scatter must be due to parameters other than age, e.g. metallicity
inhomogeneity, intrinsic variations in stellar atmospheres, or some
other not well understood or accounted for physics of stellar
evolution. This implies, of course, that a single star age, even on
the main sequence turn-off, can differ significantly to that more
appropriately determined via fitting of its coeval siblings. In this
section, we describe our procedure for fitting isochrones to determine
ages for chemically tagged groups.

\subsection{Isochrone fits}
\label{isochrone fits}
We used the Yale-Yonsei version 2 (Y$^\mathrm{2}$; \citealt{yyref}) isochrone sets
in fitting all chemically tagged groups. In order to best tune our
determinations, we generated an interpolated grid of isochrones, using
the supplied YYMIX2 Fortran code, with a resolution of 100 Myr in age
from 0 to 14 Gyr, and 0.01 dex in both [Fe/H] and [$\alpha$/Fe], each
covering the entire range of abundances in our data.

\begin{figure*}
\centering
\includegraphics[width=0.4750\linewidth]{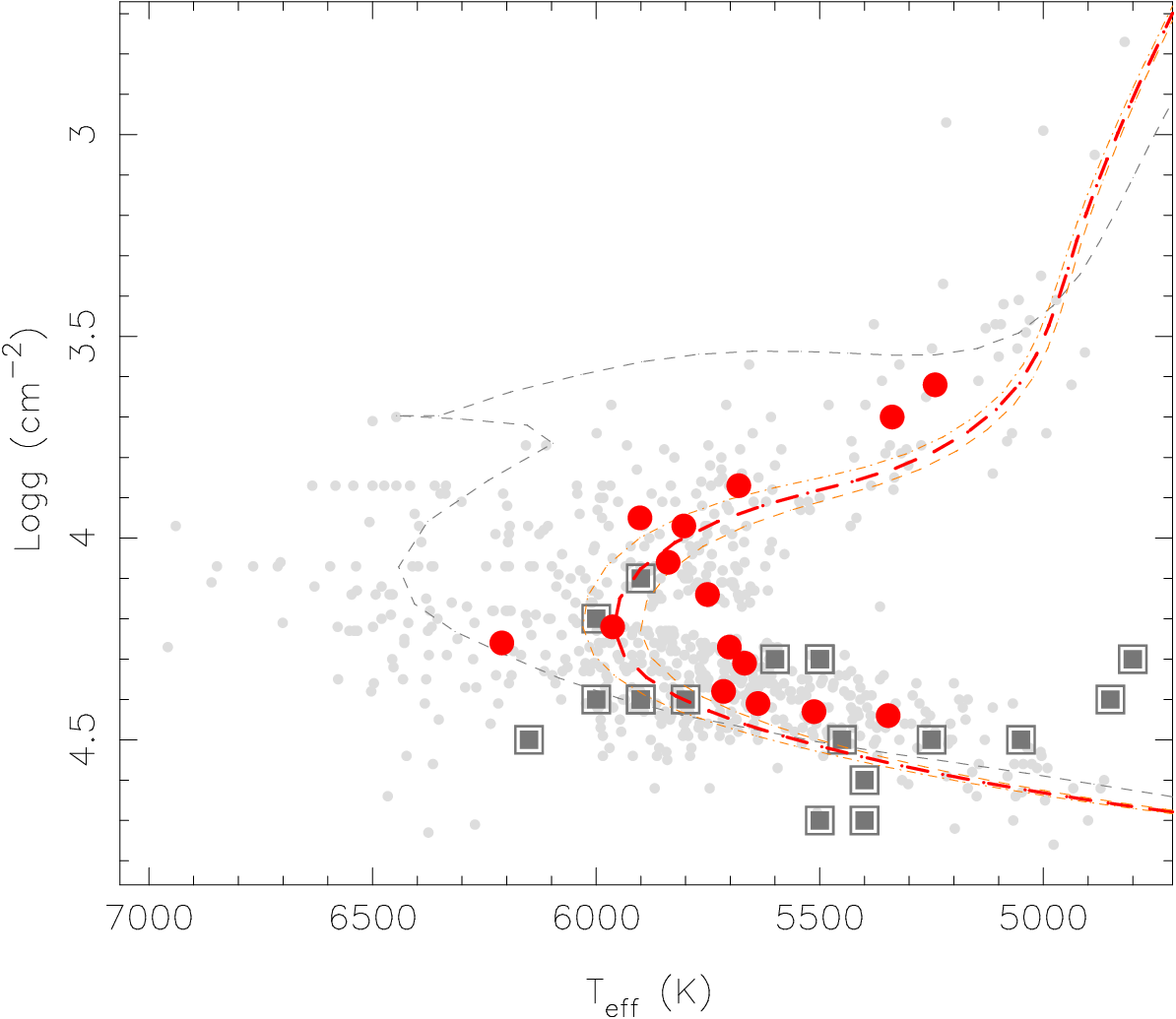}\includegraphics[width=0.4750\linewidth]{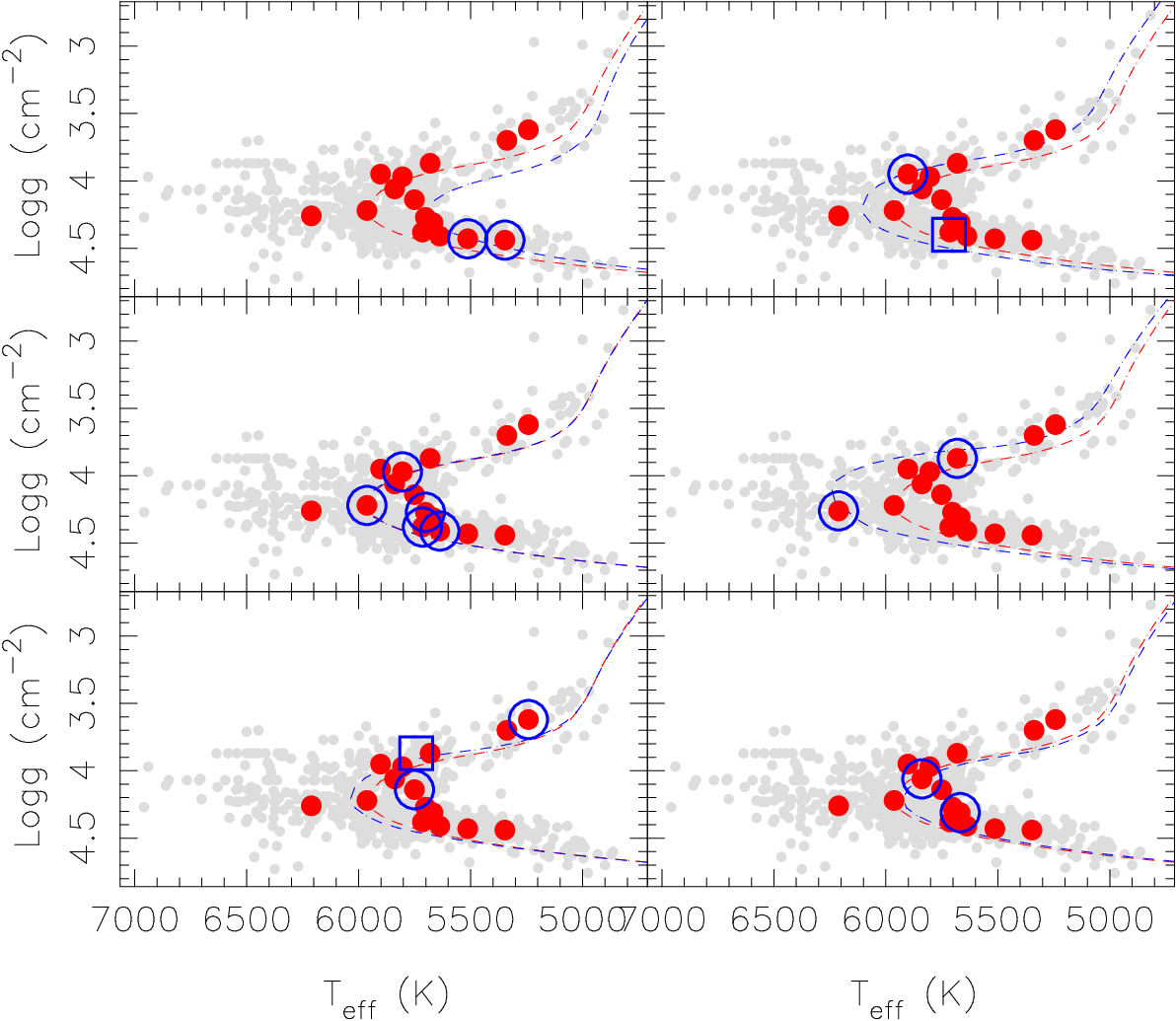}\\
\caption{Detailed view of CMD for a single group. The left hand panel shows the CMD for a group consisting of 15 members tagged to the 68\% probability threshold (red filled circles). The red dashed line shows its best-fitting isochrone, while dash-dot and dashed orange lines are best-fitting minus and plus one Gyr, respectively. Also shown are all stars in the sample for reference (light gray points), and HR1614 member stars from \citet{desilva2007a} (grey squares), along with a its best-fitting 2 Gyr isochrone (grey dashed line).  The HR1614 stars and isochrone serve to illustrate typical levels of scatter observed in conatal groups. The right hand panel shows 6 sub-groups, tagged at the 90\% probability threshold, that comprise the same group at the 68\% level. Symbols are as in the left panel, blue open symbols represent each sub-group, and the blue dashed line is the best-fit isochrone of that sub-group. Note that some subgroups have additional stars not tagged at the lower probability level (blue squares). 
\label{isofit}}
\end{figure*}

Best-fitting isochrones for each group were computed from the
resultant three dimensional grid automatically using a least squares
method. Figure \ref{isofit} shows the results of isochrone fitting for a
typical chemically tagged group. It is evident from the scatter in the
members that isochrone fits to individuals would result in
incompatible ages. However, the magnitude of scatter is actually
typical of conatal groups. The shaded square symbols show stars of
HR1614 from \citet{desilva2007a}, a well studied moving group that is
thought to be a dissolving conatal group of stars (see also
\citealt{feltzing2000}). Note the scatter of HR1614 members about its
best-fitting isochrone. Comparing that to the overall scatter of the
entire stellar sample (shown in light gray triangles) implies that, in
many cases, stellar parameters (e.g. color, magnitude, temperature,
surface gravity) do not add dimensionality to group finding using
chemical tagging. In other words, chemically tagged groups in general
will not be further refined via the position of their members in the
CMD. Inspection of CMDs in our sample confirmed this; only in a few
cases ($<$10\%) did we find obvious incompatible arrangements
(i.e., stars that appear well away from the bulk of the group, in a way
that would not satisfy stellar evolution tracks, taking into account
acceptable scatter in this plane; see e.g. (b) of Figure
\ref{grp_examples}).

\subsection{Age uncertainties}
\label{age uncertainties}
Uncertainties in our age determinations are related to uncertainties
in effective temperature and surface gravity measurements for
individual stars, and for group mean uncertainties on the
interpolation parameters [Fe/H] and [$\alpha$/Fe]. Because [$\alpha$/Fe]
was calculated by proxy, by the averaging of abundances for Ti, Mg, Si,
and Ca resulting in relatively small errors, and the effect of
modulation of this parameter on the isochrone is minimal compared to
that of [Fe/H], we ignored it in our error calculations.

\begin{figure*}
\centering
\includegraphics[width=0.4750\linewidth]{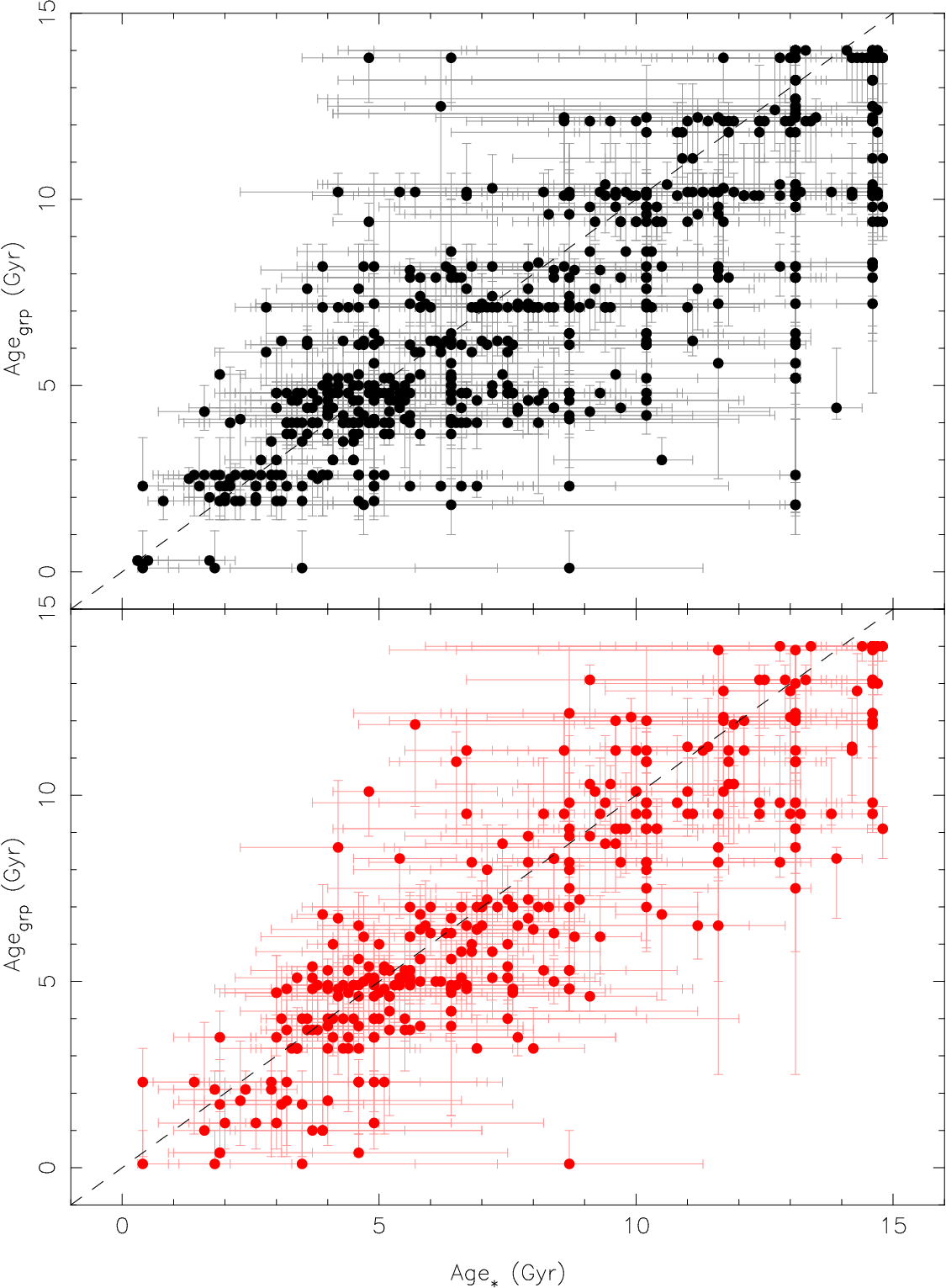}\includegraphics[width=0.4750\linewidth]{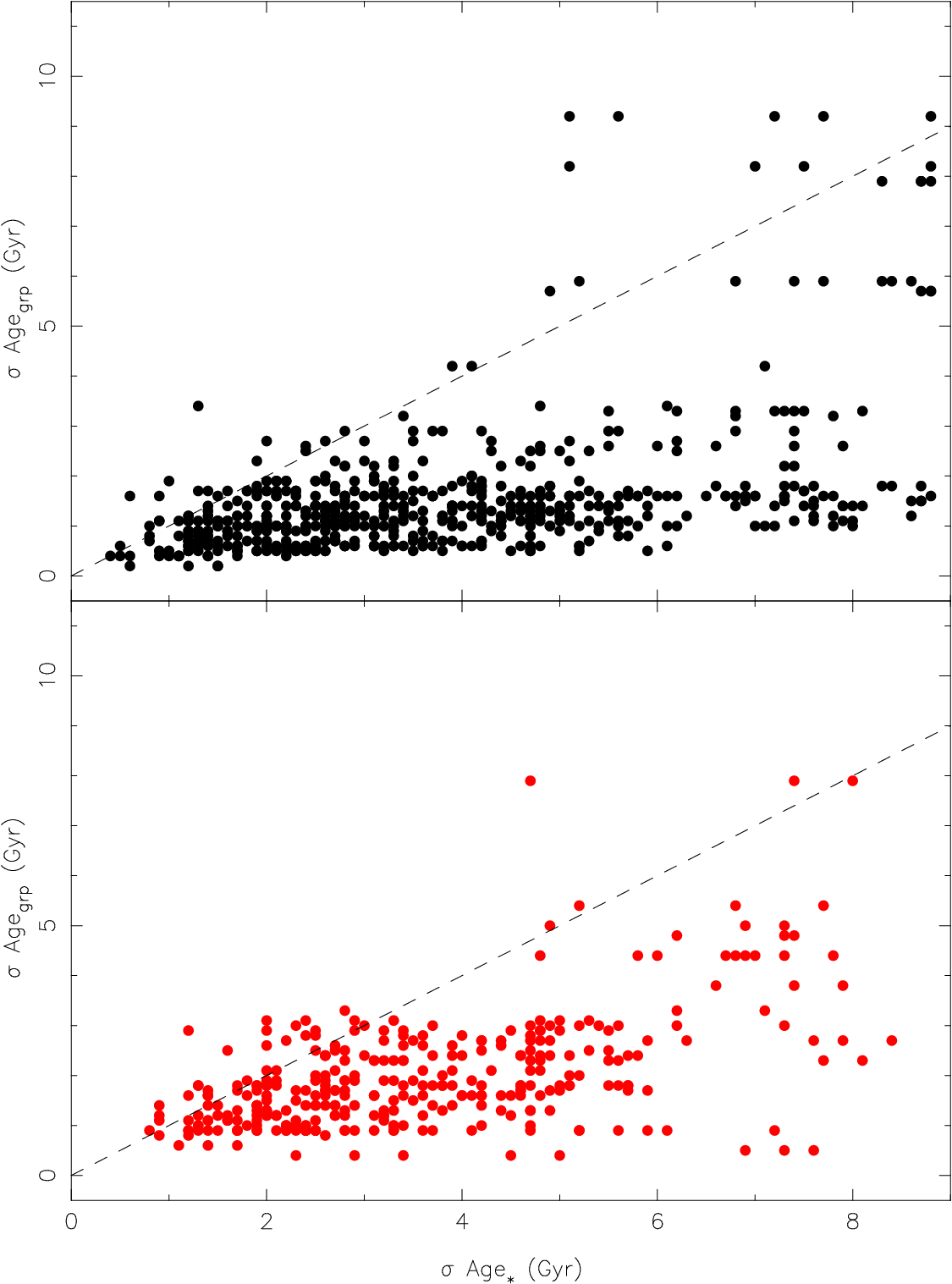}\\
\caption{Ages fit for individual stars vs those fit for clusters identified by chemical tagging. Equality is indicated by the dashed line in each panel. The top panels show members of groups at a limiting probability of 68\%, while the bottom panel shows a limiting probability of 90\%.  The left hand panels show comparison of ages while the right hand panels show comparison of the uncertainties on age. Note the difference in proportion of small groups between the two tagging thresholds, and the increase in scatter of the errors. In general, however, the group determined age errors are lower. 
\label{age_i_vs_g}}
\end{figure*}
In estimating uncertainties, we employed a Monte-Carlo approach where
each simulation iteration consisted of randomly redistributing the
stars in T$_\mathrm{eff}$ and Log g by a factor between zero and unity times their
individual uncertainties on those parameters, and re-evaluating the
best-fitting. The size of each simulation set was chosen to be 1000 and
each set was repeated three times for the cluster mean [Fe/H],
[Fe/H]-$\sigma$[Fe/H] and [Fe/H]+$\sigma$[Fe/H], resulting in a
distribution of 3000 ages for each recovered group. The 1 $\sigma$
uncertainty limits on the distribution for each group were taken as
its age errors, which are shown in the right panel of Figure
\ref{age_i_vs_g} in comparison to those from single star ages.

Of course, there must be uncertainties related to our membership
determinations. The effects of sub-sampling a coeval group of stars
and fragmentation from group finding (e.g. due to higher P$_\mathrm{lim}$) are
discussed in more detail in Section \ref{fragmentation}; however, for
the following reasons we ignore these complicated sources of error in
this discussion. Because the sub-sampling effect has a greater impact
at the measurement of a single star, our calculated uncertainties
actually represent \emph{upper limits} when making comparisons to those
from single stars. Similarly, because uncertainties on single stars
ignore intrinsic scatter in their calculations, i.e. they only take
into account uncertainties on measured parameters, they can be thought
of as representing \emph{lower limits} of the true uncertainty. Finally, as
we aim to characterize chemical tagging and study the validity of
groups derived therefrom, we must rely the assumption that our group
determinations are correct, and contamination is represented by
statistical deviations in the relations and quantities we derive.

\subsection{Fragmentation}
\label{fragmentation}
We now explore the issue of fragmentation in coeval groups linked via
chemical tagging. Fragmentation, as discussed here, is the
sub-sampling of a single chemically tagged group that is caused by
tagging at higher probabilities. In observational studies of stellar
populations we are almost always sub-sampling, or observing only a
fraction of, the whole underlying population. In chemical tagging,
effects related to this are especially important both due to the small
numbers in tagged groups, and the trade-off between high pair
probabilities within groups and the accuracy of ages determined from
them.

The right hand panel of Figure \ref{isofit} shows the fragmentation of a
single 18 member group, as tagged to the 68\% probability, when tagged
at the higher probability of 90\%. The fragmented groups in the middle
left and upper right plots exhibit noticeable discrepancies between
their best-fitting isochrones (shown in blue dashed line) and that of
the larger group, while the others are more generally consistent. The
mean and maximum absolute age differences between the fragments and
the ``parent'' group, at 1.6 Gyr and 4.7 Gyr, respectively, are less
than the same quantities compared to the ages derived for single
stars, at 2.9 Gyr and 6.4 Gyr. It is impossible in a blind chemical
tagging experiment to determine which of these fragments are truly
parts of the same population, and to what extent contamination is
affecting the particular group at lower probabilities. It is somewhat
reassuring, however, that the mean differences in age between stars
calculated from groups at the 68\% and 90\% levels, for the entire
sample, of 1.2 Gyr is close to the typical uncertainties computed as
above, and lower than the mean difference between single star ages and
those calculated for stars tagged to 68\% probability of 1.7 Gyr.

\begin{figure}
\centering
\includegraphics[width=1.0000\linewidth]{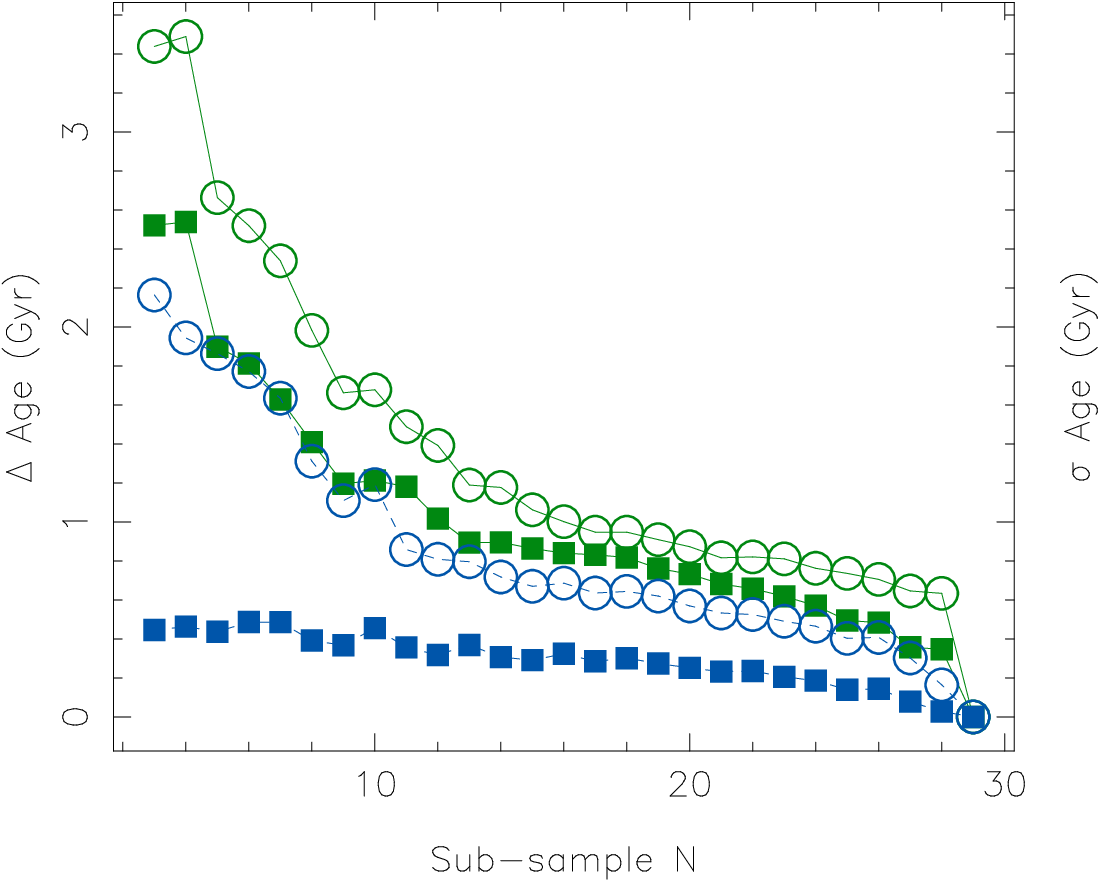}\\
\caption{Sub-sampling simulations of the Hyades. The green squares-line is the difference of the simulation mean age to the age of the Hyades, the green circles-line is the dispersion of ages in the simulated sets, as a function of sub-sample group size.  The corresponding dashed blue lines are for simulations where we guaranteed at least a single sub-giant star. The results are much as expected, with the accuracy of measurement increasing with increasing sample size. 
\label{subsample}}
\end{figure}

As seen in Figure \ref{grp_props}, a large fraction of groups have few
members. Even in the (presumably rare) case that one of these groups
has absolutely no contamination, the age we calculate for it will be
affected by the inherent sub-sampling of observational studies. The
severity of this effect is proportional to the sub-sample size, thus
these small groups present additional challenges, on top of the
uncertainties associated with chemical tagging. In an attempt to
quantify the magnitude of this effect on a chemical tagging experiment
such as this, we simulated sub-sampling of the Hyades cluster (from
\citealt{tabernero2012}) for sub-samples sized from 3 to 27 members (the
full sample having 28 stars), representing the range of coeval group
populations found in this study. Each simulation iteration selects a
random sub-set of N stars from the Hyades members, and computes the
age by best-fitting isochrone. We performed 1000 realizations of this
simulation for each size, and computed the difference of the mean age
to that of the Hyades, and the dispersion of ages in the simulated
sets, as shown in Figure \ref{subsample}. This simulation was repeated
with the condition that one of the sub-sample members is more evolved
than a dwarf star (dashed lines in Figure \ref{subsample}), which
results in improved age constraints.

The results are much as expected, with a clear trend for more accurate
ages as sub-sampling size increases. The dispersion within simulation
iterations can be thought of as a fundamental uncertainty, not related
to the accuracy of measured quantities, but rather to how well the
observed sample represents the population as a whole. With sub-giants
required as members, the accuracy (mean) substantially improves, while
the precision (dispersion) improves to a lesser degree. In addition
to providing context to the the meaning of age determinations, these
results indicate the importance of large sample sizes for chemical
tagging.

\subsection{Comparison to single star ages}
\label{comparison to single star ages}
\begin{figure}
\centering
\includegraphics[width=1.0000\linewidth]{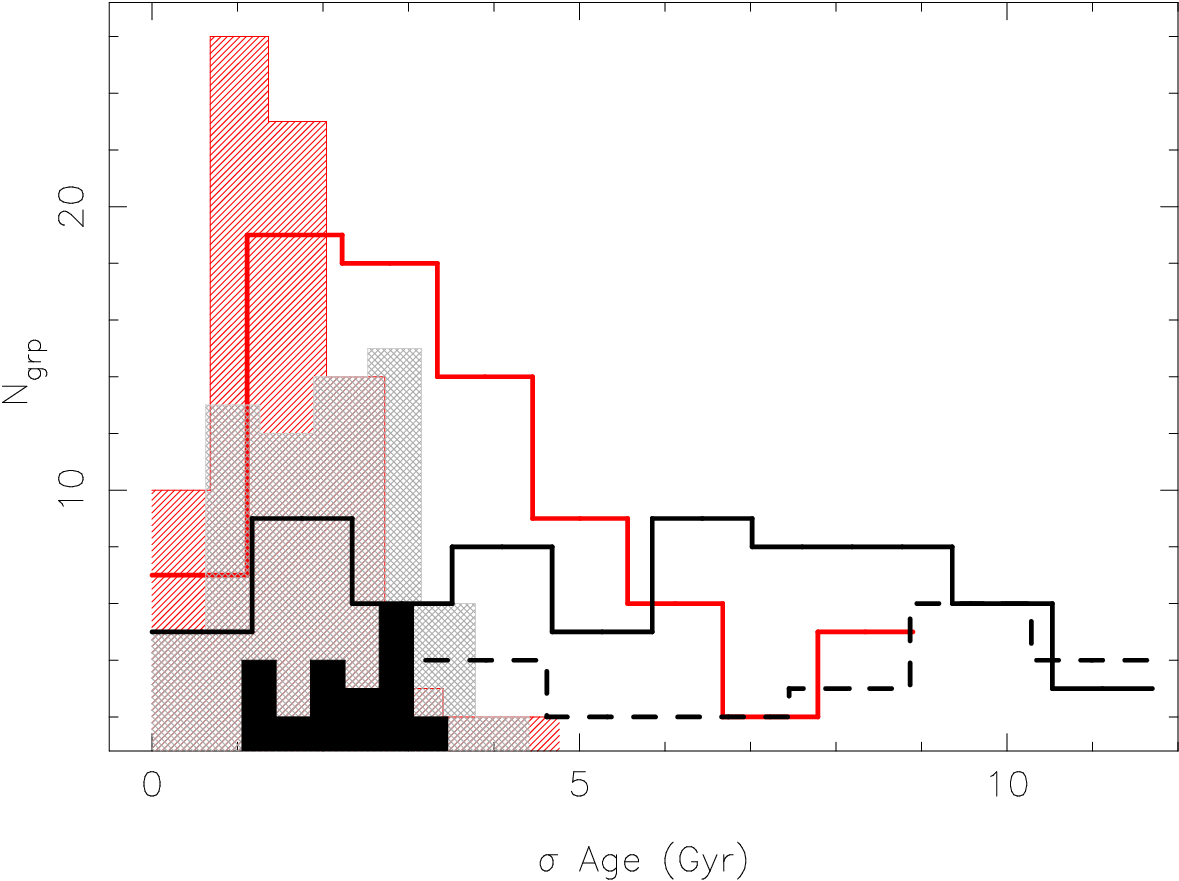}\\
\caption{The distribution of scatter of single star age determinations within chemically tagged groups. The grey and red hatched histograms are the dispersions of single star ages within groups from the 68 and 90\% pair probability thresholds with 3 or more members, respectively. The filled black histogram shows the 68\% case for groups with 10 or more members. The black and red lines show the breadth of age differences within groups for the same thresholds, while the dashed black line shows the same for 68\% groups with 10 or more members. The breadth is defined as the difference between maximum and minimum single star age within a chemically tagged group. 
\label{age_c_disp}}
\end{figure}

The isochrone fitting procedure described above was completed for all
chemically tagged groups. The results are shown in Figures
\ref{age_i_vs_g}, and \ref{age_c_disp}. The former shows a direct
comparison between single star ages and chemically tagged ages in the
left hand plot. A general agreement exists, yet there is significant
scatter and a tendency for single star ages to be larger than the
respective group ages. Departures from a one to one relationship
between age determinations are also common in our results, with a
slight tendency for turn-off and sub-giant branch stars to have
greater agreement between methods.

Figure \ref{age_c_disp} shows the scatter of single star ages within
coeval groups. Each bin in the filled histograms represents the number
of groups where the single star ages exhibit the dispersion
indicated. The line plots show the breadth of age differences in the
groups. This is defined as the difference between the largest and
smallest single star age within a group.  The mean dispersion of over
$\sim{}$2 Gyr (in the $P_{lim}=68$\% case) suggests the prospects of using
ages determined through isochrone fitting of individuals stars as an
added group finding dimension are fairly poor. Dispersions over 6 Gyr
are seen, and the breadth of disagreement extends well past 10 Gyr. If
we extend the distribution of groups found herein to a much larger
sample (e.g. a million stars) this will amount to a significant number
of coeval groups with very large dispersions in single star ages.

Though the accuracy of age determinations via chemical tagging is
difficult to quantify, being dependent on the accuracy of the group
finding, the precision can easily be compared to that of single
stars. The improvement in precision of chemically tagged ages over
single star ages is illustrated in the right panel of Figure
\ref{age_i_vs_g}, which shows the uncertainties on age derived from the
single star method compared with those from the chemically tagged
groups. The improvement can be quite substantial -- even order of
magnitude differences are seen -- and is evidently non-linear, due to
the variety of arrangements on the CMD plane afforded by multiple
stars.

\section{Age Trends in Chemically Tagged Groups}
\label{age trends in chemically tagged groups}

\subsection{Age velocity relations}
\label{age velocity relations}
The groups we have identified via chemical tagging do not unanimously
exhibit clustering in Galactic kinematic (UVW) space (as shown in
Figure \ref{grp_examples}). Given a typical cluster lifetime
(i.e. before total dissolution) of on order 10-100 Myr
(\citealt{janes1988,lada2003}), and an age range on our cluster population
from less than 1 to 14 Gyr, there is a proportionally very large span
of time for respective members to evolve dynamically within the
Galactic environment. The churning and radial migration processes
described in \citet{sellwood2002} imply that older populations would
exhibit greater velocity dispersions. The stochastic nature of the
churning process, however, would cause the relationship between
velocity and age to loosen for older epochs, unfortunately making
kinematics a poorly constrained dimension in chemical tagging group
detection.

\begin{figure*}
\centering
\includegraphics[width=1.0000\linewidth]{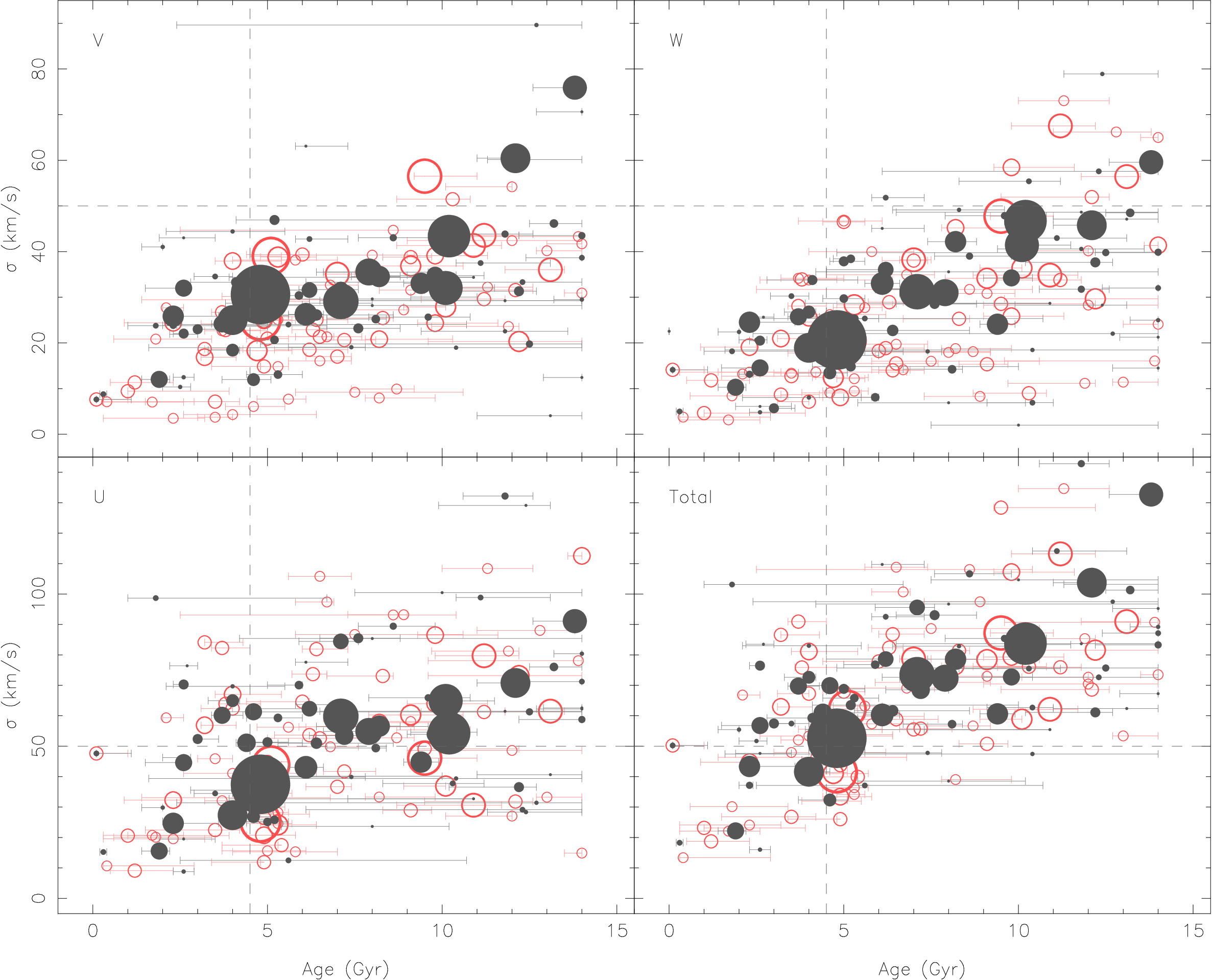}\\
\caption{Age vs internal velocity dispersions of chemically tagged groups in the U, V, W components and the total dispersion, as indicated on the plots. Filled grey circles are groups tagged to $P_{lim}=68$\% threshold, open red circles are to the 90\% threshold. The sizes of symbols represent relative sizes of groups. All groups plotted here have three or more members. The dashed axes at 4.5 Gyr and 50 km s$^\mathrm{-1}$ in each panel elucidate the difference in $\sigma$-scale of the top and bottom panels. 
\label{age_vs_disp}}
\end{figure*}

The overall age-velocity trends we observe using chemically tagged
ages suggest the validity of this picture. Figure \ref{age_vs_disp}
shows age-velocity relations (AVRs) for all three Galactic velocity
components, and the total (quadratic mean of the three
components). The vertical axis is the velocity dispersion of the
component indicated in the plot, and each point is the dispersion
between all members of a single group. There is clearly a relation
observed in both the total and W components. The V and U components
may similarly exhibit relations, but the scatter is significant,
particularly in the U component. Notice also the difference in scale
of the top two panels to the bottom two, which is highlighted by the
dashed horizontal lines at 50 km s$^\mathrm{-1}$.

The existence of a relation in total internal dispersion follows
intuitively from star formation scenarios in which clusters disperse
on short time-scales. When cluster disruption time-scales are shorter
than the dynamical time-scales, older groups would have more time
subject to Galactic churning processes than their younger
counterparts, and thus exhibit a larger velocity dispersion amongst
their constituents. One might expect the terminus of this relation at
0 Gyr to be very close to 0 km s$^\mathrm{-1}$, representing the velocity dispersion
of the parent molecular cloud at the time of star formation (typically
less than $\sim{}$1 km s$^\mathrm{-1}$), however, upon examination of Figure
\ref{age_vs_disp} this does not appear to be the case (assuming a linear
relation), being somewhere upwards of $\sim{}$30 km s$^\mathrm{-1}$. Assuming the groups
are coeval, this result may lend further to option 4 discussed in
Section \ref{results of group finding}, because the stars included in
the chemically tagged group from distinct \emph{sites} would tend to
increase the total dispersion, even at very young ages. In other
words, perhaps the AVR indicates that the groups are coeval, but not
conatal.

The age-W velocity relation, or disc heating signature, is the most
widely studied of these relations. In a dual disc (thick and thin)
scenario (e.g. see \citealt{gilmore1983}), the nominal thick disc W
dispersion at roughly $\sim{}$40 km s$^\mathrm{-1}$ is higher than the thin by about 20
km s$^\mathrm{-1}$. The trends amongst our groups are largely consistent with this
picture, showing a smooth heating signature between extremes.

A powerful distinguishing element between the two discs has been shown
to be position in the [Fe/H] and [$\alpha$/Fe] abundance plane, which
can either be represented by proxy (in the case of this study, the
average of Si, Ti, Ca and Mg), or through individual $\alpha$ elements
(\citealt{fuhrmann1998,feltzing2003,bensby2003,bensby2005,reddy2003,reddy2006,navarro2011}).
Thick disc stars are typically metal poor and $\alpha$-enhanced while
thin disc stars are metal rich and $\alpha$-normal. When stars in the
group sample are assigned to thin or thick disc based on the following
criteria (similar to the \citealt{navarro2011} criteria, but selected on
visual inspection of the [Fe/H] vs. [$\alpha$/Fe] diagram of the present
sample):

\vspace{1em}
\noindent{}Thin~disk~if~~
$\left\{\parbox{\linewidth}{\textrm{
    [Fe/H] $> -0.6$, \\
    \\
    {[$\alpha{}$/Fe]} $< -0.1 \times{}$ [Fe/H] + 0.1}}\right.$
\vspace{2em}
\\
\noindent{}Thick~disk~if~~
$\left\{\parbox{\linewidth}{\textrm{
    $-1 <$ [Fe/H] $< -0.1$, \\
    \\
    {[$\alpha$/Fe]} $> -0.1 \times$ [Fe/H] + 0.1}}\right.$
\vspace{1em}

\noindent{}we get the age distribution functions for each disc shown
in Figure \ref{disc_age_func}. Each component follows a reasonably
expected profile, with the thick disc peaking at $\sim{}$10 Gyr while
the thin is dominated by stars $\sim{}$5 Gyr old. The tails on
chemically tagged age distributions are less prominent than the same
distributions using ages derived for single stars, and the thick disc
distributions clearly exhibit differing shapes in their profiles
towards older epochs. Nevertheless, an important implication here, in
agreement with, and perhaps strengthening, conclusions of
\citet{bensby2013}, is that stellar age can act as a disc membership
discriminant; few thick disc stars should be younger than 9 Gyr, while
few thin disc stars should be older.

\subsection{Age Metallicity Relation of coeval groups}
\label{age metallicity relation of coeval groups}
Stellar elemental abundances as a function of age, the so-called
age-metallicity relation (AMR), is the fossil record of the chemical
enrichment history of the Galactic disc. This relation is fundamental
to a broader understanding of the Galaxy, however, there is not yet
agreement on its observational properties
(e.g. \citealt{rocha-pinto2000a,feltzing2001,haywood2006,soubiran2008,casagrande2011,anguiano2013a}).
The natures of these relations, if indeed they exist, remain uncertain
due to the difficulty in obtaining accurate ages for field stars, and
the difficulties in defining and observing complete samples.

\begin{figure}
\centering
\includegraphics[width=1.0000\linewidth]{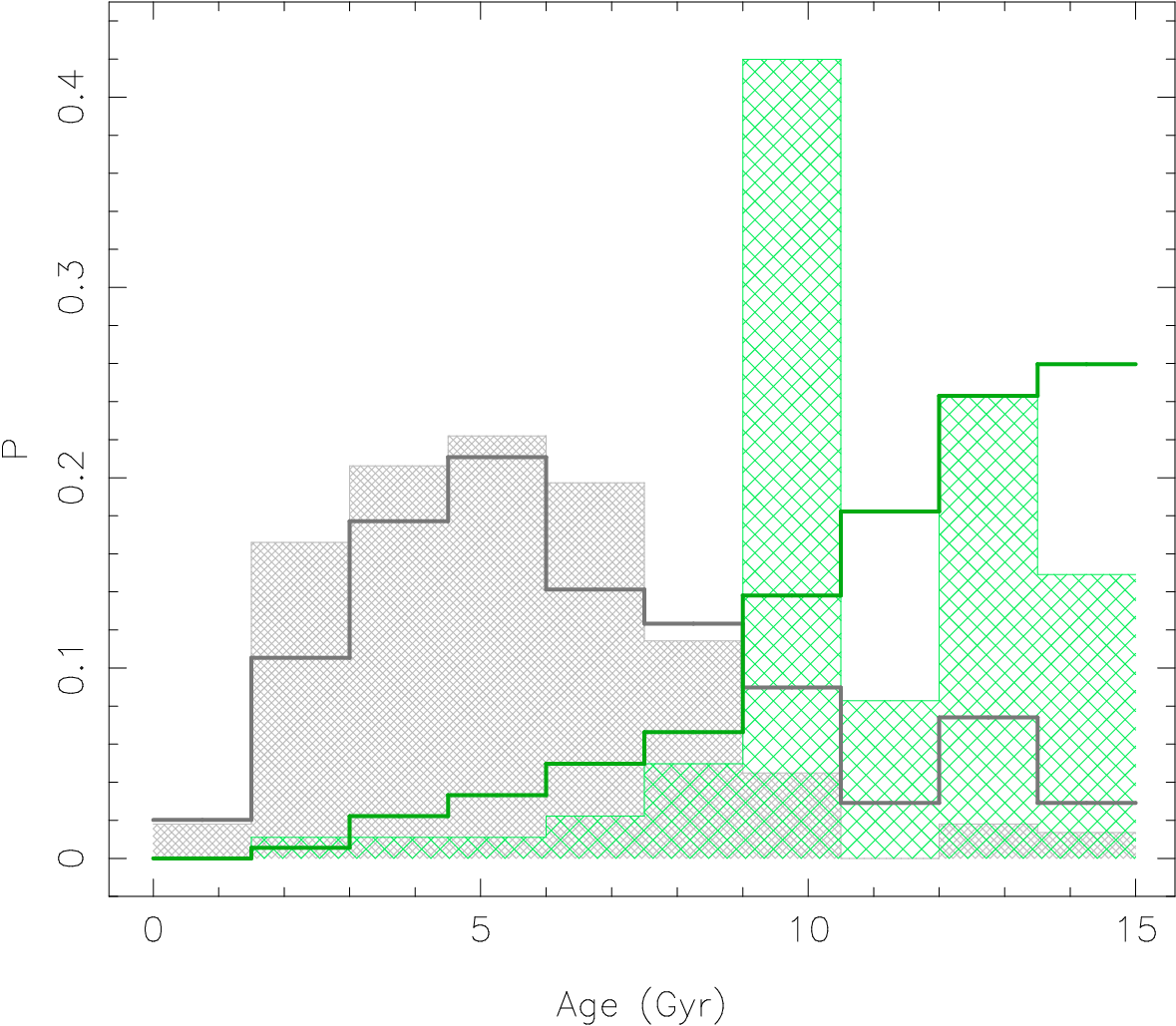}\\
\caption{Age distribution functions for the abundance selected thin (grey) and thick (green) discs. The shaded histograms are chemical tagging ages while the solid lines are single star ages -- including only stars that are also included in the filled histograms -- for reference. 
\label{disc_age_func}}
\end{figure}

\begin{figure}
\centering
\includegraphics[width=1.0000\linewidth]{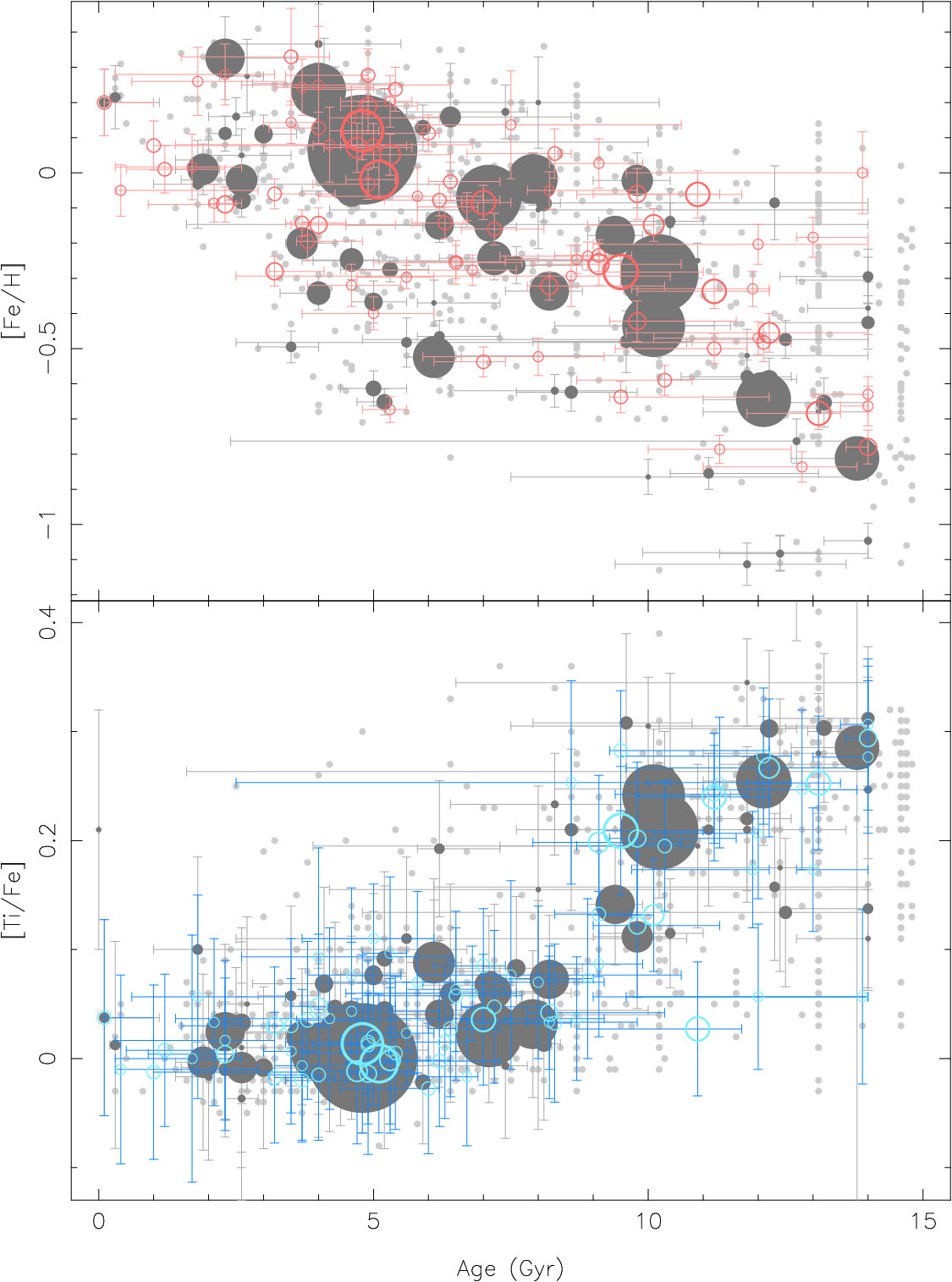}\\
\caption{Age vs metallicity for coeval groups using chemically tagged ages and mean [Fe/H] abundances. Filled grey circles are groups tagged to a $P_{lim}=68$\% threshold, red open circles are to the 90\% threshold. The size of symbols represents the relative size of groups. All groups plotted here have three or more members. The light grey points show the individual stars which are members of a 68\% tagged group with greater than three members, but with their single star age. The bottom panel is the same as the top but showing [Ti/Fe] as a function of age (here the open symbols are blue). The trends are broadly consistent, except the coeval groups appear to exhibit a tighter trend and steeper drop off in [Fe/H] beyond approximately 9 Gyrs, while the Ti abundances rapidly rise at the same cutoff. 
\label{age_vs_M}}
\end{figure}

In Figure \ref{age_vs_M}, we have plotted [Fe/H] as a function of
chemically tagged ages for coeval groups.  The relationship appears
roughly linear, while the scatter in metallicity with age, calculated
by averaging the dispersion of [Fe/H] in 2 Gyr bins, is 0.26 dex,
approximately 0.06 dex less than the scatter from single star ages
(plotted as light grey dots). The relationship trends from slightly
super-solar metallicity for the youngest group of stars, intersecting
solar level abundances at close to the Sun's age of $\sim{}$5 Gyrs,
and descending at the oldest ages to [Fe/H] of $-1$ dex. This result
is fairly consistent, if not steeper in general, than that predicted
for the solar vicinity by recent theoretical work
(e.g. \citealt{roskar2008b,minchev2013}). The scatter in metallicity
vs. age is significant enough to make comparisons with the simulations
difficult, especially since they typically show non-linearity in the
relation toward older ages; however it is worth mentioning that
\citet{roskar2008b} show, in their simulated data, that an in-situ
population in the solar vicinity, as opposed to one including radial
migration, exhibits a steeper, and tighter, AMR.  The bottom panel
  of Figure \ref{age_vs_M} is similar to the top, except showing the
  [Ti/Fe] abundances. The slope is quite shallow, but positive, for
  young stars, and consistent with single star ages. We see a knee at
  around 9 Gyr, prior to which there is a a rapid rise in $\alpha$
  abundances with respect to [Fe/H]. The age at which this abundance
  knee is observed is the same age that appears to separate the
  abundance determined thin and thick disc stars seen in Figure
  \ref{disc_age_func}.

One could argue about the presence of a knee in the [Fe/H]
distribution. If indeed present in these data, it is certainly a weak
signature. The bi-modality of the alpha abundance relation with age,
however, is unambiguous. Recently, \citet{bovy2012} suggested that the
Galaxy does not have a distinct two-component disc in terms of
scale-height, but rather a smooth distribution, and that the
[$\alpha$/Fe] vs [Fe/H] bimodality previously seen was merely a
selection effect. The W-component AVR in the top right panel of Figure
\ref{age_vs_disp}, seems to argue in favor of a smooth distribution,
given the smooth monotonic heating signature seen, though we must note
the kinematic selection of the data may preclude such analysis. The
chemical evolution is clearly not smooth, however, and the disc
appears to have two distinct components in the sense that something
triggered a change in the mode of enrichment around 9 Gyrs ago.
Given that stars of ages greater than 9 Gyrs are predominantly thick
disk, this could be indicitive of a separate star formation history
for these two populations.

\section{The Nature of Chemically Tagged Groups}
\label{the nature of chemically tagged groups}
The initial and operating assumption for much of this work has been
that chemically tagged groups represent conatal groups of stars. In
light of results from the previous sections, we revisit the
interpretations introduced in Section \ref{results of group finding}
that aim to explain the seemingly large numbers of stars in this local
sample that were tagged to groups. Four options were proposed which
included unexpectedly high contamination levels in chemically tagged
groups, open clusters as non-representative of typical Galactic star
formation events, very poor mixing efficiencies within the disc, and
finally homogeneity of chemical evolution on a Galactic scale as
opposed to a local molecular cloud scale. We reiterate that the
calibration used for determining the probability limits chemical
tagging is based on open clusters, which have been shown to not
exhibit an AMR (e.g. see \citealt{pancino2010}), and are further at odds
with our understanding of ``typical'' star formation due to the fact
that we observe intermediate age and old open clusters, which
presumably should have dispersed many millions or billions of years
prior. Thus they may not be the best calibrator for field stars
(unfortunately, there are no better calibrators at this point in
time). Adding to that, the purely chemical approach to the empirical
probability function ignores the \emph{a-priori} probability of a pair of
stars being born together in a given volume. In a Galaxy with few
clusters that remained localized, the function would be weighted
towards conatal groups, enhancing the ability to tag stars at a given
chemical difference. In a galaxy with many clusters and efficient
mixing, the function would be weighted towards distinct sites of
formation, reducing or eliminating the ability to tag, in a given
volume.

The consistency in our results with respect to the
age-metallicity-velocity (AMVR) relations indicate that option number
1, noted in Section \ref{results of group finding}, cannot be wholly
responsible. The fact that open clusters are seen to very old ages,
and the lack of an age-metallicity relation amongst them, already
indicates that they do not represent typical star formation events in
the Galaxy. Therefore option number 2 is certainly possible, yet,
again, the consistency in AMVR we find seems to indicate the group
members have a relationship beyond their chemistry. Numerical Galaxy
evolution simulations suggest that radial migration is widespread
(\citealt{roskar2008a,roskar2008b,minchev2011}), possibly even responsible
for kinematically heating the thick disc
(\citealt{loebman2011}). Observational evidence of this process, however,
is difficult to come by. Assuming this is the case would rule out
option number 3, leaving 4 as the most likely candidate. This scenario
could explain the AMVR, and would mean that a chemical signature, to
the precision we are able to currently measure it, does not represent
a single site of star formation, but rather the prevailing chemical
abundance pattern during a Galactic epoch of star
formation. Alternatively, making the assumption that chemically tagged
groups represent conatal stars, we could make the conclusion that
radial migration is not as strong a factor as simulations suggest.

In a blind chemical tagging experiment, with presently available data,
there is no robust check one can do to ensure the correct assumptions
went into linking groups. Qualifying the chemically tagged ages
against established age relations in the literature is also
problematic given the issues related to computing ages for individual
stars. However, we find encouraging the consistency of these results
with the broad generalizations of how the Galaxy, and stars within it,
evolved.

\section{Summary \&{} Conclusions}
\label{summary_and_conclusions}
We have performed the first ever blind chemical tagging experiment on
a sample of 714 stars with high-resolution abundance measurements of
12 elements. Using the methodology in \citet{mitschang2013} we linked
chemically tagged groups in two probability regimes, $P_{lim}=68$ and
$P_{lim}=90$\%, yielding 70 and 71 coeval group detections with an
average of $\sim{}$8 and $\sim{}$4 stars each, amounting to 80 and 40\% of stars in
the entire sample tagged to associations with 3 or more members,
respectively.

Several challenges present themselves. The seemingly large fraction of
stars tagged would imply weak churning efficiencies if these groups
represented discrete sites of star formation within a molecular
clouds. Yet, evidence is mounting that these mixing processes are
quite strong. Alternatively, the coeval groups may represent Galactic
epochs of star formation, in which the nuclear enrichment processes
follow the same patterns regardless of position in the Galaxy, to
within the tolerance of age measurements. The fact that open clusters
are the only source of empirical calibration on the chemical tagging
probability function is problematic, particularly for the latter
interpretation. There is no doubt, however, that a calibration of the
probability function using a large, homogeneously analyzed, open
cluster sample, e.g., that from the Gaia-ESO Survey, will help to
better understand the results of blind chemical tagging experiments.

Traditional stellar group membership criteria, i.e. those used to
determine membership for globular and open clusters, do not apply to
group finding via blind chemical tagging. Kinematic membership
criteria for open clusters require very low internal velocity
dispersions and common space motions, yet we have seen high dispersion
amongst chemically tagged groups, and the patterns do not appear
similar to Galactic moving groups. Kinematics may be loosely used as
an additional parameter to group linking through the total internal
velocity dispersion relation with age. It is unclear how useful this
will be in practice. In most cases, the scatter in the isochrone of
potential members is not enough to use those stellar parameters as a
discriminator. In our sample, less than 10\% of recovered groups
exhibited obvious outliers in the CMD plane.

Additionally, our results indicate that stellar ages do not add
another dimension to the chemical tagging group finding parameter
space. This is especially true for the most common method of computed
ages for single stars, isochrone fitting. Few of the detected coeval
groups have pre-computed (single star) ages that are compatible with
the age computed from the group. Although we cannot be definitive in
the matching of stars to coeval groups, given the relations derived
from chemically tagged ages, and the problems associated with
isochrone fitting for single stars, chemical tagging may present a
viable alternative to computing ages for stars in the samples of
upcoming large scale surveys like GALAH and Gaia-ESO.

We have shown that the results of coeval group linking, in particular
the abundance and kinematic relations with chemically tagged ages, are
consistent with modern broad understanding of the nature of their
evolution. The importance of accurate astrophysical ages cannot be
overstated. Regardless of the interpretation of the groups, whether
they represent discrete sites, or simply epochs of formation, these
ages provide a powerful diagnostic, and the consistency observed here
is encouraging looking forward to the upcoming large-scale chemical
tagging experiments.

\section{Acknowledgements}
\label{acknowledgements}
AWM gratefully acknowledges the Australian Astronomical Observatory
for a PhD top-up scholarship grant which has, and will continue to
support this work. The authors would like to thank members of the
Macquarie University and Australian Astronomical Observatory joint
Galactic archaeology group, Daniela Carollo, Valentina D'Orazi, and
Sarah Martell for helpful discussions. AWM and DBZ thank Hans-Walter
Rix at the Max-Planck Institute for Astronomy in Heidelberg, Germany
for discussion and suggestions related to the interpretation of
chemically tagged groups. DBZ acknowledges the support of Future
Fellowship FT110100743 from the Australian Research Council. T.B. was
funded by grant No. 621-2009-3911 from The Swedish Research
Counsil. S.F. was partly funded by grants No. 621-2011-5042 from The
Swedish Research Counsil.
\bibliographystyle{apj}
\bibliography{paper}
\end{document}